\documentstyle[11pt]{article}
\def\PRL #1 #2 #3{{\sl Phys. Rev. Lett.} {\bf#1} (#2) #3}
\def\NPB #1 #2 #3{{\sl Nucl. Phys.} {\bf B #1} (#2) #3}
\def\NPBFS #1 #2 #3 #4{{\sl Nucl. Phys.} {\bf B #2} [FS#1] (#3) #4}
\def\CMP #1 #2 #3{{\sl Commun. Math. Phys.} {\bf #1} (#2) #3}
\def\PRD #1 #2 #3{{\sl Phys. Rev.} {\bf D #1} (#2) #3}
\def\PLA #1 #2 #3{{\sl Phys. Lett.} {\bf #1A} (#2) #3}
\def\PLB #1 #2 #3{{\sl Phys. Lett.} {\bf B #1} (#2) #3}
\def\JMP #1 #2 #3{{\sl J. Math. Phys.} {\bf #1} (#2) #3}
\def\PTP #1 #2 #3{{\sl Prog. Theor. Phys.} {\bf #1} (#2) #3}
\def\SPTP #1 #2 #3{{\sl Suppl. Prog. Theor. Phys.} {\bf #1} (#2) #3}
\def\AoP #1 #2 #3{{\sl Ann. of Phys.} {\bf #1} (#2) #3}
\def\PNAS #1 #2 #3{{\sl Proc. Natl. Acad. Sci. USA} {\bf #1} (#2) #3}
\def\RMP #1 #2 #3{{\sl Rev. Mod. Phys.} {\bf #1} (#2) #3}
\def\PR #1 #2 #3{{\sl Phys. Reports} {\bf #1} (#2) #3}
\def\AoM #1 #2 #3{{\sl Ann. of Math.} {\bf #1} (#2) #3}
\def\UMN #1 #2 #3{{\sl Usp. Mat. Nauk} {\bf #1} (#2) #3}
\def\FAP #1 #2 #3{{\sl Funkt. Anal. Prilozheniya} {\bf #1} (#2) #3}
\def\FAaIA #1 #2 #3{{\sl Functional Analysis and Its Application}
{\bf #1} (#2) #3}
\def\BAMS #1 #2 #3{{\sl Bull. Am. Math. Soc.} {\bf #1} (#2) #3}
\def\TAMS #1 #2 #3{{\sl Trans. Am. Math. Soc.} {\bf #1} (#2) #3}
\def\InvM #1 #2 #3{{\sl Invent. Math.} {\bf #1} (#2) #3}
\def\LMP #1 #2 #3{{\sl Letters in Math. Phys.} {\bf #1} (#2) #3}
\def\IJMPA #1 #2 #3{{\sl Int. J. Mod. Phys.} {\bf A #1} (#2) #3}
\def\AdM #1 #2 #3{{\sl Advances in Math.} {\bf #1} (#2) #3}
\def\RMaP #1 #2 #3{{\sl Reports on Math. Phys.} {\bf #1} (#2) #3}
\def\IJM #1 #2 #3{{\sl Ill. J. Math.} {\bf #1} (#2) #3}
\def\APP #1 #2 #3{{\sl Acta Phys. Polon.} {\bf #1} (#2) #3}
\def\TMP #1 #2 #3{{\sl Theor. Mat. Phys.} {\bf #1} (#2) #3}
\def\JPA #1 #2 #3{{\sl J. Physics} {\bf A#1} (#2) #3}
\def\JSM #1 #2 #3{{\sl J. Soviet Math.} {\bf #1} (#2) #3}
\def\MPLA #1 #2 #3{{\sl Mod. Phys. Lett.} {\bf A #1} (#2) #3}
\def\JETP #1 #2 #3{{\sl Sov. Phys. JETP} {\bf #1} (#2) #3}
\def\JETPL #1 #2 #3{{\sl  Sov. Phys. JETP Lett.} {\bf #1} (#2) #3}
\def\PHSA #1 #2 #3{{\sl Physica} {\bf A #1} (#2) #3}
\def\CQG #1 #2 #3{{\sl Class. Quantum Grav.} {\bf #1} (#2) #3}
\def\SJNP #1 #2 #3{{\sl Sov. J. Nucl. Phys. (Yadern.Fiz.)} {\bf #1} (#2)
#3}
 \def\a{\alpha}\def\b{\beta}\def\g{\gamma}\def\d{\delta}\def\e{\epsilon}
  
 \def\s{\sigma} 
 \def\Om{\Omega} 

\newcommand{\p}[1]{(\ref{#1})}
\begin{document}
\renewcommand{\thefootnote}{\fnsymbol{footnote}}
\thispagestyle{empty}
\begin{flushright}
{\bf TUW/97-11  \\
hep-th/9707110 \\
1997, July 11 
\\ Correction, 1997 December 6,\\
Published in {\sl Phys.Lett.}{\bf B}}

\end{flushright}

\medskip
\begin{center}
{\Large\bf
A Polynomial First Order Action
for the
Dirichlet $3$--brane}

\vspace{0.3cm}

\renewcommand{\thefootnote}{\dagger} \vspace{0.2cm}

{\bf Igor A. Bandos},

\vspace{0.2cm}
{\it International Center for Theoretical Physics, \\
  34100, Trieste, Italy }
\\ and \\
{\it Institute for Theoretical Physics
\\NSC Kharkov Institute of Physics and Technology}
\\{\it 310108, Kharkov,  Ukraine}\\
e-mail:  bandos@susy-1.kipt.kharkov.ua
\\ bandos@tph32.tuwien.ac.at

and

{\bf Wolfgang Kummer,}

\vspace{0.2cm}
{\it Institut f\"{u}r Theoretische Physik,
\\ Technische Universit\"{a}t Wien, \\
Wiedner Hauptstrasse 8-10, A-1040 Wien} \\
e-mail:  wkummer@tph.tuwien.ac.at

\vspace{1.5cm}

{\bf Abstract}

\end{center}

A new first order action  for $type~IIB$ Dirichlet 3-brane is proposed.
Its form is inspired by the superfield equations of motion
obtained recently from the generalized action principle.
The action involves auxiliary symmetric spin tensor fields.
It seems promising
for a reformulation of the generalized action in a structure
 most adequate for investigating  the extrinsic geometry
of the super--3--brane, but also
for further  studies of string dualities.

\bigskip

\setcounter{page}1
\renewcommand{\thefootnote}{\arabic{footnote}} \setcounter{footnote}0

\newpage
\section{ Introduction }

Dirichlet (super)--p--branes \cite{Dbranes}--\cite{Townsend}
recently have received much attention as
objects related to
nonperturbative superstring (or M--)theory physics,
(see e.g. \cite{DbraneM}, \cite{DbraneQ}).

The original D--brane action is  of Dirac--Born--Infeld (DBI) type
\cite{Dbranes,c0}

\begin{equation}
\label{DBI}
I_{DBI} = - \int_{{\cal M}_0} d^{p+1} \xi \sqrt{-det (g_{mn} +
 {\cal F}_{mn}} ),
\end{equation}
where
$$
g_{mn}
= \partial_m X^{\underline{m}} \eta_{\underline{m}\underline{n}}
\partial_m X^{\underline{n}}
$$
is the induced metric, and the field ${\cal F}_{mn}$
are the components of the 2--form field strength
 \begin{equation}
\label{calF}
{\cal F} = dA = d\xi^m \wedge d\xi^n {\cal F}_{nm}, \qquad A=d\xi^m A_m .
\end{equation}
For simplicity we will consider  D-brane actions in
a flat background here, though their curved space generalizations
are straightforward.

The supersymmetric generalization of this action was found
recently \cite{c1}-- \cite{bt} and was used for
constructing the generalized action
\cite{bsv} for Dirichlet superbranes \cite{bst}, which allows
to obtain the superfield equations of motion for
super-D-p--branes \cite{bst}. The linearized form of such equations
was discovered previously   \cite{hs1} in the frame of the so--called
supersymmetric geometric approach \cite{bpstv,bsv1}.

Within the last months several authors
\cite{Hull,l2,Brandt}
for different reasons
adressed the problem how to avoid the appearance of the square root
in \p{DBI} by a transition to  another action which becomes an analog
of the so--called Polyakov action \cite{Pol}. Such an action,
found in fact for the first time in Ref. \cite{l1}, offers the possibility
to study rigid symmetries, duality properties and the strong coupling
limit of $D$--branes. However, the action \cite{l1},
used in \cite{Hull,l2,Brandt}, 
 is still not of polynomial type. An auxiliary world volume tensor field
 and its inverse are necessary to write down the action.

 The bosonic limit of the generalized action
for the Dirichlet super--p--branes (Dp--branes)
\cite{bst}
is already of  first order  in the field strength of the
 world volume gauge field.

However, its structure
is not completely identical
\cite{bst}
to the
 one for superstrings and $type~I$ superbranes
\cite{bsv,bsv1}. If we consider the pure bosonic limit (the
so--called
moving frame or Lorentz harmonic formulation of the bosonic branes, see
\cite{bzp,bpstv,baku}) for the latter
in a flat  background,
it is of first order in all the dynamical fields. The bosonic limit of the
generalized action of the Dirichlet superbranes,
being of the first order in the auxiliary
gauge field strength ${\cal F}$,
remains  of $p+1$-th order in the derivative
of the embedding coordinate functions $X^{\underline{m}}(\xi^{m})$
($\underline{m}= 0,...,9; ~~m=0,...,p$).

Thus the natural question to ask is whether it is possible to write
the generalized action for super--D-p--branes and, hence,
the moving frame action for D--branes in  first order form.

The purpose of this note is to give an affirmative answer.
 We present this new first order form of the action
for a particular, but especially  interesting, case of a $type ~IIB$
D3--brane.
The basical  requirement for such an action has been to reproduce the
relation between induced and intrinsic vielbeins, most adequate
for the description of the embedding of superspaces
\cite{hs2,bst}. It is gratifying that this in a straightforward way leads
 even to a polynomial form of the action.

Our result provides a
basis for an improved reformulation
of the generalized action  which, in turn,  enables
further  studies of
the extrinsic geometry of D--brane world volume superspace. It also
 seems to be a convenient new element for
 the investigation of duality problems.

In addition our result can be regarded as a preliminary step for obtaining
a generalized action for the $M$--theory super--5--brane
which is to be still beset by some problems
\cite{bpst}.

In section 2 we summarize the information about the bosonic limit
of the generalized action for $D=10$ D-p--branes. The adequate world volume
vielbein of section 3 is the basis for the construction of our new action
in terms of adequate auxiliary fields in Section 4.
In the Conclusion we also return to
problems of the M--theory 5--brane.

\section{ Moving frame action and embedding equations for $D=10~~$
Dirichlet p--branes.}

The  moving frame action for the bosonic $D_p$-brane \cite{bst}
  in flat $D=10$
space--time reads
\footnote{In this section for the simplicity we drop the dependence
on all background fields, including the 2--form $B_2$ and the dilaton $\phi$.}
 \begin{equation}\label{mfac}
S_D =
\int_{{\cal M}^{p+1}}
({\cal L}^0_{p+1} +
{\cal L}^1_{p+1})
\end{equation}
 \begin{equation}\label{mfac0}
{\cal L}^0_{p+1}=
{1 \over (p+1)!} E^{a_0} \wedge ... \wedge
 E^{a_p} \epsilon_{a_0 ... a_p}
\sqrt{-det (\eta_{ab} + F_{ab})}
\end{equation}
 \begin{equation}\label{mfac1}
{\cal L}^1_{p+1} = Q_{p-1} \wedge
[ dA -{1\over 2} E^b \wedge E^a F_{ab}]\Big).
\end{equation}
In \p{mfac0} $Q_{p-1}$ is the Lagrange Multiplier $(p-1)$--form,  
\begin{equation}
\label{Ea}
E^a = dX^{\underline{m}}(\xi) u_{\underline{m}}^a (\xi)
\end{equation}
is the pull--back of the $(p-1)$ components of a target space vielbein form
(
$\underline{m}=0,...,9;$ $\underline{a}=0,...,9;$
$a=0,...,p;$ $i=1,...,9-p$)
\begin{equation}
\label{Eua}
E^{\underline{a}}= dX^{\underline{m}}
u_{\underline{m}}^{~\underline{a}}
= (E^a, E^i)
\end{equation}
which is related to the holonomic vielbein $dX^{\underline{m}}$ by a
Lorentz rotation parametrized by the matrix
\begin{equation}
\label{harm}
(u_{\underline{m}}^{~\underline{a}}) =
(u_{\underline{m}}^{a},
u_{\underline{m}}^{i}) \qquad \in \qquad SO(1,9) \qquad
\Leftrightarrow \qquad
u_{\underline{m}}^{~\underline{a}}
\eta^{\underline{m}\underline{n}}
u_{\underline{n}}^{~\underline{b}} =
\eta^{\underline{a}\underline{b}}
\end{equation}
This Lorentz rotation is chosen to adapt the vielbein \p{Eua}
to the world volume of the (Dirichlet) $p$--brane in the sense that
only $(p+1)$ components $E^a$ of \p{Eua} enter into the action.
Thus the $u_{\underline{m}}^{\underline{a}}$ depend on the world volume
coordinates $\xi^m$ and shall be regarded as (auxiliary) field variables
(Lorentz harmonic variables, see \cite{gikos,sok} and refs. in \cite{bpstv}).
They are to be varied to obtain the equations of motion on equal footing with
$X^{\underline{m}}$, taking into account
the conditions \p{harm}. To avoid the introduction
of the  constraints \p{harm} with Lagrange multipliers into the action,
we can restrict the variations of
$u_{\underline{m}}^{~\underline{a}}$
to the space
 isomorphic to the Lie algebra of the Lorentz group
\cite{bzp}.
Such
variations shall be just infinitesimal Lorentz rotations

\begin{equation}
\label{harmvar}
\d u_{\underline{m}}^{~\underline{a}} =
u_{\underline{m}}^{~\underline{b}}
{\cal O}_{\underline{b}}^{~\underline{a}} , \qquad
{\cal O}^{\underline{b}\underline{a}} = -
{\cal O}^{\underline{a}\underline{b}}.
\end{equation}
This
defines  ''admissible'' derivatives
of the harmonic variables
\begin{equation}
\label{dharm}
d u_{\underline{m}}^{~\underline{a}} =
u_{\underline{m}}^{~\underline{b}}
\Om_{\underline{b}}^{~\underline{a}}  \qquad
\end{equation}
as well, where
\begin{equation}\label{Cf}
\Om^{\underline{a}\underline{b}} = -
\Om^{\underline{b}\underline{a}} =
\left(
\matrix{ \Om^{ab} & \Om^{aj} \cr
       - \Om^{bi} & \Om^{ij} \cr}
        \right)
= u^{~\underline{a}}_{\underline{m}} d u^{\underline{b}\underline{m}}
\end{equation}
is an $so(1,D-1)$ valued Cartan $1$--form
\footnote{The parameters of admissible variations ${\cal O}$ can be considered as
contractions of the Cartan forms \p{Cf} with the variation symbol
$\d $, $~~~{\cal O}^{\underline{a}\underline{b}} = i_\d
\Om^{\underline{a}\underline{b}} $.}.

Clearly, the definition \p{Cf} of the Cartan forms can be regarded as the statement
that they are trivial $SO(1,9)$ connections. The condition of vanishing
for the corresponding $SO(1,9)$ curvature (equivalent to \p{Cf})
coinsides with the Maurer--Cartan equations
\begin{equation}\label{MC}
d\Omega^{\underline a\underline b} -
\Omega ^{\underline a}_{~\underline c}
\wedge \Omega^{\underline c \underline b}
= 0,  \qquad  \Leftrightarrow \qquad
 \cases{
{\cal D}\Omega ^{ai} \equiv
d\Omega ^{ai} -
\Omega ^{a}_{~b} \wedge \Omega ^{bi} +
\Omega ^{aj} \wedge \Omega ^{ji} = 0,  \cr
R^{ab}=
d\Omega ^{ab} -
\Omega ^{a}_{~c} \wedge \Omega ^{cb} =
\Omega ^{ai} \wedge \Omega^{bi},  \cr
R^{ij} =
d\Omega ^{ij} +
\Omega ^{ij^\prime} \wedge \Omega^{j^\prime j} = -
\Omega ^{ai} \wedge \Omega^{~j}_{a}, \cr }
\end{equation}
Eqs. \p{MC} for the forms $\Om^{ai}, \Om^{ab}, \Om^{ij}$ \p{Cf}
pulled back to the world volume
give rise to the Peterson--Codazzi, Gauss and Ricci
equations of  surface theory \cite{Ei}.

Splitting the expression for the admissible variations
in an $SO(1,p)\times SO(9-p)$ invariant way (cf. \p{Eua} \p{harm}, \p{Cf}),
we get in particular
\begin{equation}
\label{deltaua}
\d u_{\underline{m}}^{a} =
u_{\underline{m}}^{b} {\cal O}_b^{~a} -
 u_{\underline{m}}^{i} {\cal O}^{ia}
=
u_{\underline{m}}^{b} i_\d \Om_b^{~a} +
 u_{\underline{m}}^{i} i_\d \Om^{ai} .
\end{equation}
Eq. \p{deltaua} can be regarded as the contraction of the
expression for admissible
derivatives
\begin{equation}
\label{dua}
d u_{\underline{m}}^{a} =
 u_{\underline{m}}^{b}  \Om_b^{~a} +
 u_{\underline{m}}^{i}  \Om^{ai}  , \qquad   \Leftrightarrow \qquad
D u_{\underline{m}}^{a} \equiv
d u_{\underline{m}}^{a} -
 u_{\underline{m}}^{b} \Om_b^{~a} =
 u_{\underline{m}}^{i} i_\d \Om^{ai}  . \qquad
\end{equation}
For completeness let us present the analogous expression for the
derivatives of the harmonic  $u_{\underline{m}}^{i}$ :
\begin{equation}
\label{dui}
d u_{\underline{m}}^{i} =
 - u_{\underline{m}}^{j}  \Om^{ji} +
 u_{\underline{m}a}  \Om^{ai}
\qquad   \Leftrightarrow \qquad
D u_{\underline{m}}^{i} \equiv
d u_{\underline{m}}^{a} +  u_{\underline{m}}^{j}  \Om^{ji} =
 u_{\underline{m}a}  \Om^{ai}
\end{equation}

It turns out \cite{bst} that the parameter
${\cal O}^{ba}=i_\d \Om_{ba}$ \p{deltaua} of the $SO(1,p)$ subgroup of the
$SO(1,9)$ can be identified with the world volume local
Lorentz symmetry of the D-p-brane model and yields no
independent equations of motion (Noether identity), while
the variation ${\cal O}^{ai}=i_\d \Om_{ai}$ provides us with
the equation  (see \cite{bpstv,bsv,baku,bst})
\begin{equation}
\label{Ei=0}
E^i \equiv dX^{\underline{m}} u_{\underline{m}}^{~i}  = 0 .
\end{equation}
In Eq. \p{Ei=0} the statement that the vielbein $E^{\underline{a}}$ is adapted to
the embedding is thus realized in a  concrete way.

The additional assumption that the pull-backs of the remaining vielbein forms
are linearly independent (which is a nondegeneracy condition on the
embedding)
means that, in general,
\begin{equation}
\label{Ea=ebmba}
E^a \equiv dX^{\underline{m}} u_{\underline{m}}^{~a}  =
e^b m_b^{~a},
\end{equation}
  where the matrix  $m_b^{~a}$ is supposed to be nondegenerate
$ det(m_b^{~a})\not= 0 $.
For any choice of the matrix $m$,  for the induced metric
\begin{equation}
\label{gind}
g_{mn} \equiv
\partial_m X^{\underline{m}}
\eta_{\underline{m}\underline{n}}
\partial_m X^{\underline{m}}
\equiv
E_m^{\underline{a}} \eta_{\underline{a}\underline{b}}
E_n^{\underline{b}} =
E_m^{~a} \eta_{ab}
E_n^{b}
\end{equation}
holds, where Eq. \p{Ei=0} was taken into account.
This freedom to choose a convenient matrix $m$ will be used below. It is an
essential ingredient of our argument.

In terms of \p{Ea=ebmba}
the Nambu--Goto volume form is rewritten as
  \begin{equation}
\label{NGvol}
d^{p+1}\xi \sqrt{-det(g_{mn})} =
d^{p+1}\xi det(E^a_{m}) =
{ 1 \over (p+1)! } \e_{a_0...a_p} E^{a_0} \wedge ... \wedge E^{a_p}.
\end{equation}

The action \p{mfac} includes the
auxiliary antisymmetric tensor superfield $F_{ab}$,
and the $(p-1)$--form
Lagrange multiplier $Q_{p+1}$
in addition to the world volume
gauge field $A_m$.
The Lagrange multiplier $Q_{p+1}$
produces the equation
\begin{equation}\label{calF=F}
{\cal F} \equiv dA = {1 \over 2} E^a \wedge E^b F_{ba}
\end{equation}
which can be solved algebraically with respect to the
auxiliary field $F_{ab}$ expressing it in terms of the
field strength ${\cal F}_{mn}$ \p{calF}.
On the other hand, the variation of the action with respect
to the auxiliary field $F_{ab}$ yields
\begin{equation}
\label{dS/dF}
Q_{p-1} \wedge E^b \wedge E^a ={ \sqrt{det(\eta +F)}\over (p+1)!}
E^{a_0} \wedge ... \wedge E^{a_p}
\e_{a_0 ... a_p} (\eta + F)^{-1~[ba]}
\end{equation}
whose solution for
$Q_{p-1}$ is
  \begin{equation}
\label{Q=}
Q_{p+1} ={ \sqrt{det(\eta +F)}\over 2(p-1)!}
E^{a_1} \wedge ... \wedge E^{a_{p-1}}
\e_{a_1 ... a_{p-1}bc} (\eta + F)^{-1~bc} .
\end{equation}

Thus $Q_{p-1}$ does not contain propagating degrees of freedom.

The dynamical equations appear in the moving frame formulation
as follows. One varies the action with respect to the embedding
functions $X^{\underline{m}}$ and the gauge fields $A_m$
\begin{equation}
\label{dS/dX}
d\Big(({1 \over p!} E^{a_1} \wedge ... \wedge E^{a_p}
\epsilon_{a_1... a_p a}
\sqrt{-det (\eta_{ab} + F_{ab})} -
 Q_{p-1} \wedge E^b  F_{ab}) u_{\underline{m}}^{a} \Big) = 0,
\end{equation}
\begin{equation}
\label{dS/dA}
d Q_{p-1} =0
\end{equation}
 and  uses the algebraic equations \p{Ei=0}, \p{calF=F},
\p{Q=}
to exclude auxiliary variables from Eqs. \p{dS/dX}, \p{dS/dA}. Replacing
the induced vielbein by the induced metric
\p{gind} one then reproduces the equations following from \p{DBI}.
In this way the classical equivalence of the
moving frame formulation of the D-p-branes \p{mfac} with the standard
Born--Infeld--like one \p{DBI} can be proved.

To see this equivalence at the level of the action functionals,
 the algebraic equation is used
to remove the auxiliary field from the functional
 \p{mfac}. The second term \p{mfac1} vanishes as a result of Eq. \p{calF=F}.
Henceforth
the
auxiliary field $F_{ab}$ has to be replaced
by the field strength
${\cal F}_{ab} = {\cal F}_{mn} {\cal E}^m_a {\cal E}^n_b$
( ${\cal E}^m_a E^a_n = \d^m_n$ )
in the first term \p{mfac0}.
The  square root multiplier in
the first term of the functional  \p{mfac} may be written as
${\sqrt{-det(g_{mn}+{\cal F}_{mn})} \over \sqrt{-det(g_{mn})}}$,
 where $g_{mn}$ is the induced metric \p{gind}. Then, using
the consequence \p{NGvol} of the algebraic equation \p{Ei=0},
one  gets the standard DBI--like action \p{DBI}
(see \cite{bst} for more details as well as for the supersymmetric
generalization).

\section{The adequate world volume vielbein.}

The  moving frame action for strings and for $type~I$ p--branes
\cite{bzp,bpstv} can be written as \cite{bpst}
\footnote{Eq. \p{mfacI} superficially coincides with
\p{NGvol},
however it has the nontrivial property to produce Eq. \p{Ei=0} which is necessary
for the identification with Nambu--Goto action
\p{NGvol}.}
 \begin{equation}
\label{mfacI}
S_I =
\int_{{\cal M}^{p+1}}
{1 \over (p+1)!} E^{a_0} \wedge ... \wedge
 E^{a_p} \epsilon_{a_0 ... a_p} .
\end{equation}
However, its original form was of the first order
with respect to the $X$ variable
\cite{bzp,bpstv}
 \begin{equation}\label{mfacI1}
S_I^\prime  =
\int_{{\cal M}^{p+1}}
\Big({1 \over p!} E^{a_0} \wedge e^{a_1} \wedge ... \wedge
 e^{a_p} \epsilon_{a_0 ... a_p}
- {1 \over (p+1)!} e^{a_0} \wedge e^{a_1} \wedge ... \wedge
 e^{a_p} \epsilon_{a_0 ... a_p} \Big)
\end{equation}
where $e^a= d\xi^m e_m^a(\xi )$ is the (auxiliary) world volume
vielbein 1--form field.

The action functional \p{mfacI1}  in addition to the equation \p{Ei=0}
produces Eq. \p{Ea=ebmba} with unit matrix $m$
\begin{equation}
\label{Ea=ea}
E^a \equiv dX^{\underline{m}} u_{\underline{m}}^{~a}  = e^{a} .
\end{equation}
 Such an identification of the intrinsic world volume vielbein with the induced
one is indeed natural for  $type~I$ extended objects (superbranes).
For the supersymmetric case  it results
in the standard expression (constraint) for the dimension 1
bosonic world volume torsion superform
\cite{bsv}
\begin{equation}
\label{Ta}
T_{\a q~\b p}{}^a \propto \g^a_{\a \b} C_{qp} ,
\end{equation}
and the fermionic equations acquire the standard form with
vanishing gamma trace of a spin 3/2 superfield
\begin{equation}
\label{eqmferm}
\g^a_{\b \a} \psi^{a \a q} =0 .
\end{equation}
The linearized approximation of \p{eqmferm}
is given essentially by
the vector derivative of the target space Grassmann coordinate superfield
$\Theta^{\underline{\mu}}$
\begin{equation}
\label{dTheta}
\psi_a^{\a q} \approx \partial_a \Theta^{\underline{\mu}}
v_{\underline{\mu}}^{~\a q} .
\end{equation}
In Eqs. \p{Ta}, \p{eqmferm}, \p{dTheta} $\a $ and $q$ are $SO(1,p)$
and $SO(D-p-1)$ spinor indices carried
by a world volume Grassmann covariant derivative $D_{\a q}$
and by a world volume Grassmann vielbein form $e^{\a q}$.
 $\g^a_{\a \b}$ are $SO(1,p)$ sigma matrices and
$C_{qp}$ is second rank invariant spin--tensor of the $SO(D-p-1)$ group.
$v_{\underline{\mu}}^{~\a q}$ denotes the
 ${SO(1,D-1) \over SO(1,p)\times SO(D-p-1)}$
spinor Lorentz harmonic variable (see \cite{bpstv} and refs. therein).

To reach the same standard structure for the case of $D=10$ $type~ II$
super--Dp--branes \cite{hs1,bst}
(and the M-theory 5-brane \cite{hs2}), the
invertible matrix $m$ \p{Ea=ebmba} must be chosen to be
depend on the field strength ${\cal F}$ of the
world volume vector field (or of the selfdual worldvolume antisymmetric
tensor field of the five--brane \cite{hs2}) \cite{bst}.

Thus for the case of a $type~IIB$ 3--brane which we will consider below,
an adequate choice of the matrix $m$ is given by \cite{bst}
\begin{equation}
\label{m=}
m_a^{~b} = \d_a^{~b} + {1 \over z(F)\bar{z}(F)} Tr(\tilde{\s}_a F \s^b \bar{F}) .
\end{equation}
In the trace of \p{m=} $F,~\bar{F}$ are spinor representations for the
selfdual and the
anti--selfdual part of the tensor $F_{ab}$
\begin{equation}\label{f1}
F_{\a\b} = F_{\b\a} =
{i\over 4}   F^{ab} (\s_a \tilde{\s}_b)_{\a\b} ,
\qquad
\bar{F}_{\dot{\a}\dot{\b}} =
\bar{F}_{\dot{\b}\dot{\a}} =
-{i\over 4}   F^{ab} (\tilde{\s}_a
\s_b)_{\dot{\b}\dot{\a}} ,
\end{equation}
\begin{equation}\label{f2}
F_{\a\dot{\a}}^{~\b\dot{\b}} \equiv
F_a^{~b} \s^a_{\a\dot{\a}} \tilde{\s}_b^{\dot{\b}\b} =
2 \d_\a^{~\b} \bar{F}_{\dot{\a}}^{~\dot{\b}} +
2 \d_{\dot{\a}}^{~\dot{\b}}
F_\a^{~\b} , \qquad
{}^{*}F_{\a\dot{\a}}^{~\b\dot{\b}} =
2 \d_\a^{~\b} \bar{F}_{\dot{\a}}^{~\dot{\b}} -
2 \d_{\dot{\a}}^{~\dot{\b}}
F_\a^{~\b} , \qquad
\end{equation}
and the scalar factors $z$ and its complex conjugate
 $\bar{z}$ (called ${1 \over 2b_{\pm}}$
in \cite{bst})
are expressed through the
$F_{ab}$ tensor as
\begin{equation}
\label{zz}
z = {1  \over 2}
\Big(1 + {i\over 8} \e^{abcd} F_{ab} F_{cd}
+ \sqrt{-det(\eta + F)}\Big)
, \qquad
\end{equation}


The action for the D3--brane
can be represented as
 \begin{equation}\label{mfac'}
S_{D3}^{\prime} =
\int_{{\cal M}^{1+3}}
({\cal L}^{0\prime }_{4} +
{\cal L}^1_{4}) ,
\end{equation}
where
 \begin{equation}\label{mfac01'}
{\cal L}^{0\prime}_{4}=
det(m)
\sqrt{-det (\eta_{ab} + F_{ab})}
\Big({1 \over 3 \cdot 3!} E^{a^\prime} m^{-1}{}^{~a}_{a^\prime}
\wedge e^{b} \wedge e^c \wedge   e^{d} \epsilon_{abcd} -
{1 \over 3 \cdot 4!} e^{a} \wedge e^{b} \wedge e^c \wedge
 e^{d} \epsilon_{abcd} \Big)
\end{equation}
 can be obtained from \p{mfacI1} if one replaces $e^a$ by
$e^b m_b^{~a}$, includes an overall multiplier $\sqrt{-det(\eta + F)}$
(see \p{mfac0})
and uses the identities
$$
\e_{abcd}
m_{a^\prime}^{~a}
m^{b}_{b^\prime}m^{c}_{c^\prime}m^{d}_{d^\prime} =
\e_{a^\prime b^\prime c^\prime d^\prime} det(m), \qquad
\e_{abcd}
m^{b}_{b^\prime}m^{c}_{c^\prime}m^{d}_{d^\prime} =
det(m)  m^{-1}{}^{~a^\prime}_a
\e_{a^\prime b^\prime c^\prime d^\prime} , \qquad
$$
The second term in the functional \p{mfac'}
 \begin{equation}\label{mfac11}
{\cal L}^1_{4} =
Q_{2} \wedge
[
e^{-{\phi \over 2}}
(dA - B_{(2)}) -{1\over 2} E^b \wedge E^a F_{ab}]\Big),
\end{equation}
corresponds to  \p{mfac1}, but (in contrast to \p{mfac1}) with the dependence
on dilaton $
\phi = \phi (X(\xi))$
and
NS--NS 2--form
background field
$B_2 = {1\over 2} dX^{\underline{m}} \wedge dX^{\underline{n}}
B_{\underline{n}\underline{m}} (X)
$
restored.

However, the drawback of such an action is the complicated form of the
dependence on the antisymmetric tensor field
$F_{ab}$.
This indicates that the latter is not an appropriate auxiliary field for
the problem under consideration.

\section{Adequate auxiliary variables and a new first order form
of the $D_3$--brane action}

The superfield (embedding) equations \cite{bpstv}
for type $II$ super--$D_3$--brane
\cite{hs1,bst}
include
the  symmetric spin tensor (super)fields
$h_{\a\b},~\bar{h}_{\dot{\a}\dot{\b}}$. They are
 expressed in terms of anti--selfdual  and selfdual
components $F_{\a\b}$ and $\bar{F}_{\dot{\a}\dot{\b}}$ of
the antisymmetric tensor $F_{ab}$
by
\begin{equation}\label{h(f)}
h_{\a\b} = {1 \over z(F)}  F^{\a\b},
\qquad
\bar{h}_{\dot{\a}\dot{\b}} = { 1 \over \bar{z}(F)} \bar{F}_{\dot{\a}\dot{\b}} ,
\end{equation}
where the  functions $z$, $\bar{z}$ are defined in Eq. \p{zz}.
The expression \p{m=} for the matrix $m$
simplifies when written in the terms of these
spin tensors

\begin{equation}\label{m=hh}
m_a^{~b} = \d_a^{~b} + {1\over 2} Sp(\tilde{\s}_a h \s^b \bar{h}) ,
\qquad    \Leftrightarrow \qquad
m_{\a \dot{\a}}^{~\b \dot{\b}}
\equiv
m_a^{~b} \s^a_{\a \dot{\a}} \tilde{\s}_b^{~\b \dot{\b}} =
2 (
\d_\a^{~\b}  \d_{\dot{\a}}^{~\dot{\b}} +
h_\a^{~\b}  \bar{h}_{\dot{\a}}^{~\dot{\b}}).
\end{equation}
This provides  us with (indirect) evidence that precisely these spin tensors
$h_{\a\b}$, $\bar{h}_{\dot{\a}\dot{\b}}$ are adequate auxiliary variables
for the first order (super-)$D3$--brane action.

If this is true, then  all we need to do is
to express the tensor $F_{ab}$ in terms of the
spin--tensors $h$ and $\bar{h}$ by solving Eq. \p{h(f)}.

From Eqs. \p{zz} we have
\begin{equation}
\label{zz1}
z+\bar{z} = 1+ \sqrt{-det(\eta + F)} , \qquad
z- \bar{z} =
{i\over 8} \e^{abcd} F_{ab} F_{cd}
\end{equation}
whereas from Eqs. \p{f1} and \p{h(f)} we obtain
\footnote{An even more general identity
$F_{ac} {}^{*}F^{cb} = {1 \over 2} \d_a^b (z^2h^2 -\bar{z}^2\bar{h}^2)$
can be proved using the spinor representation \p{f2}.}
\begin{equation}
\label{eFF}
{i \over 4} \e^{abcd} F_{ab} F_{cd} \equiv
{1\over 2}
F_{ab} {}^{*}F^{ba}
= z^2h^2 -\bar{z}^2\bar{h}^2 , \qquad
{1\over 2} F_{ab} F^{ba}
= - z^2h^2 -\bar{z}^2\bar{h}^2 , \qquad
\end{equation}
with the abbreviations
\begin{equation}
\label{h2h2}
h^2 \equiv h_{\a\b} h^{\a\b} , \qquad
\bar{h}^2 \equiv
\bar{h}_{\dot{\a}\dot{\b}}
\bar{h}^{\dot{\a}\dot{\b}} .
\end{equation}
These relations and the identity
$$
-det(\eta +F)\equiv 1 - {1 \over 2} F_{ab}F^{ba} -
({1\over 8} \e^{abcd} F_{ab} F_{cd})^2
$$
can be used to write the product
of $z$ with $\bar{z}$ \p{zz} as
 \begin{equation}
\label{zz2}
4z\bar{z} = \left( 1+ \sqrt{-det(\eta +F)}\right)^2 +
\left( {1\over 8} \e^{abcd} F_{ab} F_{cd} \right)^2
= 2(z + \bar{z}) + z^2 h^2 + \bar{z}^2 \bar{h}^2 .
\end{equation}
From the second Eq.
\p{zz1}
together with
\p{eFF} and replacing the first Eq.
\p{zz1}
by \p{zz2} we arrive at
 \begin{equation}
\label{zz2'}
z - \bar{z} = {1 \over 2} (\bar{z} ^2 \bar{h}^2 - z^2 h^2) , \qquad
z +\bar{z} =
2 z\bar{z} -
{1 \over 2} (\bar{z} ^2 \bar{h}^2 + z^2 h^2) .
, \qquad
\end{equation}
 The sum and the difference of these equations are homogeneous in
$z$ and $\bar{z}$, respectively. Thus for nonvanishing
$z$ (as implied by Eq. \p{zz1})
we can extract a system of
linear equations for $z$ and $\bar{z}$
\begin{equation}
\label{zzlin}
1 = \bar{z} -
{1 \over 2} h^2 z , \qquad 1 = z -
{1 \over 2}\bar{h}^2 \bar{z} ,
\end{equation}
with the  solution
\begin{equation}
\label{z=}
z = { 1 + {1\over 2} \bar{h}^2 \over 1- {1\over 4} h^2 \bar{h}^2} ,
\qquad
\bar{z} = { 1 + {1\over 2}h^2 \over 1- {1\over 4} h^2 \bar{h}^2} .
\end{equation}

Substituting \p{z=} into \p{h(f)} we obtain the  expression for the
selfdual and anti--selfdual parts
\p{f1}
of
$F_{ab}$
and, hence, for the whole
tensor $F_{ab}$ from \p{f2}.
The expression for the DBI--like square root can be obtained directly from
Eqs. \p{z=} and \p{zz1}:
\begin{equation}
\label{DBI=}
\sqrt{-det(\eta + F)} = {(1+{h^2 \over 2})(1+ {\bar{h}^2 \over 2})
\over
1-{h^2\bar{h}^2 \over 4} }
\end{equation}

The inverse matrix $m^{-1}$ and the determinant $det(m)$ are
\begin{equation}
\label{m-1=}
m^{-1}{}_a{}^b=
\Big( \d_a^{~b} - {1\over 2} Sp(\tilde{\s}_a h \s^b \bar{h})\Big)
{1\over \sqrt{det(m)}} ,
\end{equation}
\begin{equation}
\label{det(m)=}
det(m)= \left( 1-{h^2\bar{h}^2 \over 4} \right)^2 .
\end{equation}
Eq. \p{m-1=} can be obtained directly from the spinor representation for the
matrix $m$ \p{m=hh} while the simplest way to obtain \p{det(m)=}
is to use  a special gauge,
where only two of the components of the tensor $F_{ab}$
\begin{equation}
\label{gauge}
 F_{01} = f_{+} , \qquad F_{34} = f_{-}
\end{equation}
are nonvanishing  (gauges of such type  were used in \cite{c0,schw51}).
One can verify that in this gauge $$z^2h^2= {1 \over 2} (f_- + i f_+)^2 , \qquad
\bar{z}^2 \bar{h}^2 = {1 \over 2} (f_- - i f_+)^2$$ and
($I_2$ is $2 \times 2$ unit)
$$
m_a^{~b} = \left( \matrix{
(1 + {f_-^2 + f_+^2 \over 2z\bar{z}}) I_2
& 0 \cr
0 &
-(1 - {f_-^2 + f_+^2 \over 2z\bar{z}}) I_2
\cr } \right)
$$
Eq. \p{det(m)=} can be easily obtained from these expressions.

Substituting  \p{DBI=}, \p{m-1=}, \p{det(m)=},
into \p{mfac'}, \p{mfac01'},\p{mfac11} we arrive at
 \begin{equation}\label{mfac''}
S_{D_3}^{\prime} =
\int_{{\cal M}^{1+3}}
({\cal L}^{0\prime }_{4} +
{\cal L}^{1\prime }_{4}) ,
\end{equation}

 \begin{equation}\label{mfac01}
{\cal L}^{0\prime}_{4}=
{\Big(1+ {h^2 \over 2}\Big)
\Big(1+ {\bar{h}^2 \over 2}\Big) \over 3 \cdot 3!}
\Big[ E^{a^\prime}
\Big(\d^{~a}_{a^\prime} -
 {1\over 2}\tilde{\s}_{a^\prime}^{\dot{\b}\b}
h_{\b}^{~\a} \s^{a}_{\a\dot{\a}} \bar{h}^{\dot{\a}}_{~\dot{\b}}\Big)
\wedge e^{a} \wedge e^c \wedge   e^{d} \epsilon_{abcd} -
\end{equation}
$$
- {1 \over 2}
\Big(1- {h^2\bar{h}^2 \over 4}\Big)
e^{a} \wedge e^{b} \wedge e^c \wedge
 e^{d} \epsilon_{abcd} \Big],
$$

 \begin{equation}\label{mfac21}
{\cal L}^{1\prime}_{4} = Q_{2}' \wedge
\Big[\Big(1-{h^2\bar{h}^2 \over 4}\Big) (dA - B_{(2)}) - 
{1\over 4} E^{\alpha\dot{\alpha}} \wedge 
E_{\beta\dot{\beta}}\left(
\d_\a^{~\b} \Big(1+{h^2 \over 2}\Big) \bar{h}_{\dot{\alpha}}^{~\dot{\beta}}  
+ \d_{\dot{\alpha}}^{~\dot{\beta}}  
\Big(1+{\bar{h}^2 \over 2}\Big) h_\alpha^{~\beta}
   \right)\Big], 
\end{equation}
where, in addition to  \p{h2h2} the bispinor representation for the
vielbein indices
$
E^{\a \dot{\a}} = E^a \tilde{\s}_a^{\dot{\a}\a} $
are used. We recall  that
$
h_{\a\b} = h_{\b\a} , ~
\bar{h}_{\dot{a}\dot{\b}} =
\bar{h}_{\dot{\b}\dot{a}}
$
are auxiliary symmetric spin tensor fields replacing $F_{ab}$.

The functional \p{mfac''}, \p{mfac01}, \p{mfac21} is our result for the first order action
for the $type~IIB$ $D3$--brane.

\section{Conclusion and discussion}

In this note we present a new first order form of the
action functional for the Dirichlet $3$--brane. It is inspired by
 the superfield equations of the $type~IIB$ super--$D_3$--brane
obtained in \cite{bst} from the generalized action
and possesses a polynomial structure in the
 auxiliary spin--tensor fields
$h_{\a\b}$ and $\bar{h}_{\dot{\a}\dot{\b}}$ which assume the place of the initial
antisymmetric tensor $F_{ab}$.
It is remarkable that the superfield counterparts of these objects appear
in the basic superfield equations (fermionic rhetropic conditions \cite{bsv})
of the super--$D3$--brane model \cite{hs2,bst}.

This action can be used as a basis for the reformulation of the
generalized action principle for the
super--$D3$--brane in a simpler and more transparent manner, similar to the one
originally proposed for superstrings and type $I$ superbranes in
\cite{bsv}.

The new polynomial  first order action also seems very promising
for further studies of  dualities following the
arguments  presented in \cite{Hull,Townsend}.

An interesting problem for further investigations is
the generalization of our result to
 the  case of $type~ II$ Dp--branes with $p>3$.
Of course, then we cannot apply the
spinor calculus in the manner which was most
convenient for the 3--brane model.  The analog of the special gauge
\p{gauge} should be used instead.

An  object similar to the spin tensors
$h_{\a\b}, ~\bar{h}_{\dot{\a}\dot{\b}}$
appears in the superfield equations \cite{hs2} for M--theory 5--brane \cite{m5}.
In that case it is a symmetric  spin--tensor field
$h_{\hat{\a}\hat{\b}} \equiv \g^{\hat{a}\hat{b}\hat{c}}_{\hat{\a}\hat{\b}}
h_{\hat{a}\hat{b}\hat{c}}~~$
($\hat{\a},\hat{\b}=1,...,4;~~$
$\hat{a},\hat{b}, \hat{c} = 0,...,5$ ) which provides a spinor
representation for the $d=6$ selfdual antisymmetric tensor
$h_{\hat{a}\hat{b}\hat{c}} = {1 \over 3!}
\e_{\hat{a}\hat{b}\hat{c}\hat{d}\hat{e}\hat{f}} h^{\hat{d}\hat{e}\hat{f}}$.

On the other hand, the original action for M--theory super--5--brane
\cite{blnpst,schw52} as well as the moving frame action described shortly
 in \cite{bpst} contains the auxiliary scalar field \cite{pst51} and
a field  strength of the second rank antisymmetric world volume tensor field.
This field satisfies a nonlinear generalization of the selfduality conditions
\cite{schw51} on the mass shell (in fact this selfduality  only appears
after gauge fixing of a special symmetry).
As it was proved in \cite{hsw,5brequiv}, the
 component equations following from the action \cite{blnpst,schw52}
 coincide with the ones having been obtained from the superfield embedding
equations  \cite{hs2}.
In this way the  (ordinary) selfdual tensor
field $h_{abc}$ was expressed in terms of the auxiliary scalar and the
field strength of
the world volume 2--form gauge field.

However, to complete the proof of the equivalence of the superfield approach
\cite{hs2} (based on the embedding equations without reference to
any action) and the component approach based on the action
\cite{blnpst,schw52,bpst},
it is necessary to lift the moving frame reformulation \cite{bpst} of the
super--5--brane action
\cite{blnpst,schw52}  to a generalized action which should produce the
superfield equations.
Such a program was completed for the
superstring, for $type~I$ super--p--branes \cite{bsv} and $type~II$ Dirichlet
superbranes \cite{bst}. However, for the case of the 5--brane
its realization has encountered a problem \cite{bpst} related to
the specific way the
auxiliary scalar field is present in the action
(see \cite{blnpst1} for more details).

One of the possibilities to overcome such a difficulty consists in searching
for another form of the super--5--brane action involving
a spin--tensor representation $h_{\hat{\a}\hat{\b}}$ of the
(linearly) selfdual tensor field $h_{\hat{a}\hat{b}\hat{c}}$
instead of a specific combination of the auxiliary scalar
field and the field strength of the world volume two form.
Our present study demonstrates that a similar program at least
can be realized for the
Dirichlet 3--brane. In this sense this note can be regarded as
  a preliminary  study  of the possibility to find a   reformulation of
the 5--brane action, although we are well aware of additional problems
which are bound to appear due to the more complicated structure
of the 5--brane theory.

\newpage

\centerline{\large \bf Acknowledgements}

The authors  thank D. Sorokin and 
 M. Tonin for interest in this work and useful
discussions.
One of the authors (I.B.) is grateful to
Prof. M. Virasoro for hospitality at the International Center for Theoretical Physics
(Trieste, Italy), where part of this work was done.
He also acknowledges partial support from the INTAS Grants  {\bf N93-127-ext,
N93-633-ext} and from the
Austrian Science Foundation (Fonds zur F\"orderung der
wissenschaftlichen Forschung) project {\bf P-10.221-PHY}.

\end{document}